\documentclass[multphys,vecphys,sechang]{svmult}

\usepackage{makeidx}     
\usepackage{graphicx}    
\usepackage{multicol}    

\makeindex             

\begin{document}

\title*{A Geometric Determination of the \\Distance to SN 1987A and the LMC}
\titlerunning{Distance   to SN 1987A and the LMC} 
\author{NINO PANAGIA\inst{1,2}}
\institute{$^1$Space Telescope Science Institute, 3700 San Martin Drive,
Baltimore, MD\\ 21218, USA; \texttt{panagia@stsci.edu}\\
$^2$Affiliated  with the Space Telescope Division of the European Space 
Agency, ESTEC, Noordwijk, Netherlands}
%
%
\maketitle

\begin{abstract}
Using the definitive reductions of the IUE light curves by 
\cite{Sonneborn97} and an extensive set of \emph{HST} images of
SN~1987A  we have repeated and improved our original analysis
\cite{Panagia91}  to derive a better determination of the distance to
the supernova. 
In this way we have obtained an absolute size of the ring $R_\mathrm{abs}  =
(6.23 \pm 0.08) \times 10^{17}$\,cm and an angular size  $R''  =  808
\pm 17$\,mas, which give a distance to the supernova $d(SN1987A) =
51.4 \pm 1.2$\,kpc and a distance modulus  $(m-M)_{SN1987A} = 18.55 \pm
0.05$. Allowing for a displacement of SN~1987A position relative to
the LMC center, the distance to the barycenter of the Large Magellanic
Cloud is also estimated to be $d(LMC)=51.7\pm1.3$\,kpc, which
corresponds to a distance modulus of $(m-M)_{LMC} = 18.56 \pm 0.05$. 
\end{abstract}

\section{Introduction\label{Sec.1}} 

Cepheid variables are possibly the most reliable, and certainly the
most widely used secondary distance indicators to measure distances up
to several tens of Mpc. Because of this they play a crucial role in the
determination of the cosmological distance scale (for a review see the
proceedings of the STScI Symposium \emph{The Extragalactic Distance
Scale}, \cite{Livio97}). On the other hand, the calibration of Cepheids
as  distance indicators is based on the study of Cepheid variables in
the  LMC and, therefore, determining the distance to the Large
Magellanic  Cloud is a fundamental step in establishing a cosmological
distance  scale because the zero point of the Cepheid calibration
relies  crucially on the calibration of the LMC distance. 

Various methods have been employed to measure the distance to the LMC,
with various degrees of success and/or accuracy (e.g.\, 
\cite{Madore91}). All methods, however, are indirect in that they all
depend on the calibration of other distance indicators, and, therefore,
have only a statistical value. Moreover, different distance indicators
appear to give discordant results that are not compatible with each
other, thus making the distance issue very slippery. 

The presence of the famous circumstellar ring around SN~1987A has 
provided a unique opportunity to determine the distance to the LMC 
\emph{directly} by using a purely geometric method: it consists in
measuring  the angular size of the ring from high resolution images and
comparing it to the absolute size as estimated from the evolution of
emission lines produced by the ring ionized gas (see e.g.\,
\cite{Panagia91,Dwek92,Gould95}).

\begin{figure}
\centering
\includegraphics[height=7.5cm]{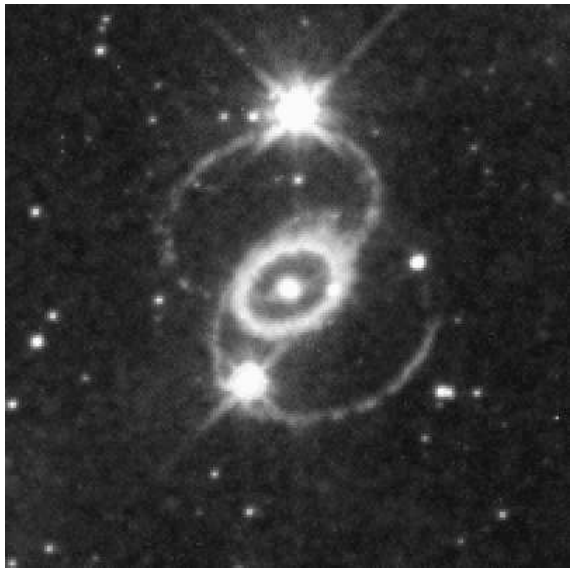}
\caption{An $8''\times8''$ region centered on SN~1987A as
observed on September~24, 1994, with the \emph{HST-WFPC2}  in an [OIII]
5007\,\AA\ filter.  In addition to the supernova, this figure shows
clearly the presence of the three circumstellar rings, a brighter
equatorial ring and two fainter, larger rings that are loosely aligned
along the polar axis. \label{Fig.1}} 
\end{figure} 

In 1991 Panagia {\it et al.}~\cite{Panagia91} estimated the distance to
SN~1987A ($51.2\pm3.1~kpc$) from a comparison of the angular size of
the inner circumstellar ring as measured with the {\it HST-FOC} in
August 1990 \cite{Jakobsen91}, with the ring absolute size as
determined from the peaks of its UV emission line light curves. 

More recently, Gould re-examined the problem adopting an infinitely narrow
ring geometry and retaining Panagia {\it et al.}~\cite{Panagia91}
assumption of an exponential law for the line emissivity
\cite{Gould95,Gould98}.  Thus, using the same data as in Panagia {\it
et al.}~\cite{Panagia91} and Sonneborn {\it et al.}~\cite{Sonneborn97},
respectively, but including only the NIV] and NIII] light curves, and
adopting the {\it average} [OIII] ring size as measured by Plait {\it
et al.}~\cite{Plait95} over the period August 1990 - May 1993, Gould
concluded that the distance to SN 1987A be less than $47~ kpc$. 

The new reductions of the IUE spectra, done by Sonneborn {\it et al.}~
\cite{Sonneborn97}, have produced more accurate and reliable light
curves. Therefore, we have decided to repeat our analysis using the new
data set and including a more accurate and realistic estimate of the
ring angular size obtained from the study of an extensive set of HST
images of SN~1987A. Here, we present a brief outline of our analysis and
the main results of our study. A complete account of this work will be
presented in a forthcoming paper \cite{Panagia03}.

\section{The Angular Size of the Ring\label{Sec.2}}

The inner circumstellar ring is clearly extended with a HPW of about
1/7 its radius (e.g. 
\cite{Jakobsen91,Panagia03}). The finite width of the ring makes the 
definition of an average size a very delicate one, which, if done 
improperly, may introduce errors as large as, say, half the HPW, 
i.e.\ as much as 7\% or more. 

Also, to derive the distance to SN\,1987A one should compare \textbf{the 
absolute ring size}, determined from the light curves \textbf{of twice
ionized N and C and three times ionized N}, \textbf{with the angular size
of the ring} as measured \textbf{at the time of the peak} for radiation
emitted by ions of \textbf{comparable ionization stages}.

\begin{figure}
\centering
\includegraphics[height=7.5cm]{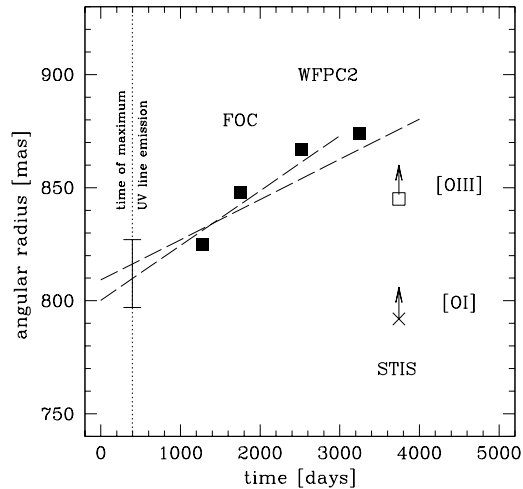}
\caption{The surface brightness averaged radius of the ring as measured
in \emph{FOC} and \emph{WFPC2} images taken with a narrow band [OIII]
filter, and the recent measurements made with the STIS in the light of
the [OIII] 5007\,\AA\ and [OI] 6300\,\AA\ lines \cite{Sonneborn98}. 
The two straight lines are the best fits to the \emph{FOC}+\emph{WFPC2}
points only, and to the  \emph{FOC}+\emph{WFPC2}+\emph{STIS} [OIII]
sizes, respectively. The error bar represents the range of possible
values of  the radius at the time of the UV maximum.\label{Fig.2}}
\end{figure}

While the images obtained with \emph{HST} in the [OIII] line filter
(see e.g.\ Fig.~\ref{Fig.1}) satisfy the second requirement ({\it
comparable ions}), they fail to satisfy the first one ({\it comparable
epochs}). On the other hand, the analysis of both \emph{FOC}  and
\emph{WFPC2} images have revealed that the ring angular size, as
estimated from  [OIII] images, appears to increase with time 
(\cite{Plait95,Panagia03}; see Fig.~\ref{Fig.2}) while the one derived
from H$\beta$ and [NII] images remains constant in time and virtually
coincides with the size measured in the earliest [OIII] image. This is
the effect of both cooling and recombination of the OIII ion, that
cause the [OIII] 5007\,\AA\ line intensity to decline more quickly at
the inner edge of the ring where the density is believed to be higher.
An experimental confirmation of this effect is provided by HST-STIS
imaging-spectroscopy of SN~1987A, obtained in April 1997
\cite{Sonneborn98},   that has shown an appreciably smaller ring size
in the [OI] 6300\,\AA\ line than it is in  the [OIII] 5007\,\AA\ line.

Therefore, the \textbf{best value  of the ring angular size} to compare
with the absolute size determined from the UV lines is an
extrapolation of the observed sizes, as measured with \emph{HST} in the
[OIII] 5007\,\AA\ filter, back to the epoch of maximum UV line emission
(approximately 400 days after the explosion, i.e.\ around early
April 1988; cf.\ Fig.~\ref{Fig.3}). In this way we obtain: 
$$R''=808\pm17\,\mathrm{mas}\,.$$

\section{The Absolute Size of the Ring\label{Sec.3}}

It has been shown \cite{Panagia91,Dwek92,Gould95} that under the
assumption of an infinitely  narrow width the \textbf{absolute radius
of the ring} can be derived from  measurements of the onset time of the
UV line emission, $t_{\circ}$, and the time of maximum  UV line
emission $t_\mathrm{max}$ because they correspond to the times when 
the near side and the far side of the ring start shining as a result of
the ionization due to the initial UV flash from the supernova
explosion. A simple geometric argument gives 
$$R=c(t_{\circ}+t_\mathrm{max})/2\,.$$

\begin{figure}
\centering
\includegraphics[height=10cm]{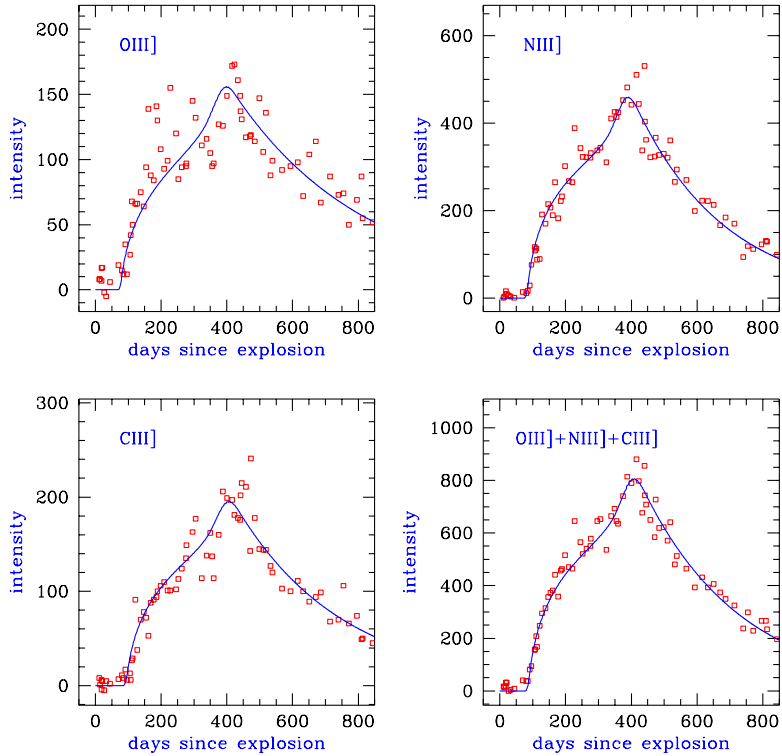}
\caption{The observed intensities (in units of
$10^{-15}$\,erg\,cm$^{-2}$\,s$^{-1}$) of the OIII] 1666\,\AA, NIII]
1750\,\AA, and CIII] 1909\,\AA\ lines and their straight sum are
presented and compared to their best-fit model light
curves.\label{Fig.3}} 
\end{figure}

As mentioned before, one has to measure the absolute size for the same
emitting ion for which one can measure the angular size. In addition, 
one has also to take into account that the ring is clearly extended 
with a width $\delta R \simeq R/7$.
Therefore, we have limited our analysis to the UV light curves of
twice ionized ions, namely OIII, NIII and CIII, and we have compared
them to theoretical light curves computed under the following
assumptions: 
\begin{itemize}
\item The ring is circular and has a gaussian width with HPW of 14\% the 
ring radius.
\item The intrinsic emission of each ion decays exponentially with 
time.
\item The free parameters are the radius and the inclination
angle of the ring, the specific emissivity at time $t=0$ and the
decay time of each line. 
\end{itemize}
The best fits to the light curves for the OIII] 1666\,\AA, NIII] 1750\,\AA,
and CIII] 1909\,\AA\ lines are shown in Fig.~\ref{Fig.3}.

We also show the composite light curve, sum of the OIII], NIII] and
CIII] line intensities, and its best model fit: we note that the
scatter in the composite light curve is greatly reduced relative to
the three light curves, indicating that most of the fluctuation is
actually noise.  The individual determinations of the absolute radius
fall in the range 230 to 248 light-days, or 6.0 to $6.4 \times
10^{17}$\,cm, with uncertainties of about 4\% for OIII] and CIII], and
slightly above 1\% for NIII], resulting in an average value of 
$$R = (6.23\pm0.08)\times 10^{17}\,\mathrm{cm}\,.$$

\section{Distance Determination\label{Sec.4}}

Comparing the absolute size of the circumstellar ring to its angular 
size, we derive a \textbf{distance to SN~1987A} 
$$\mathbf{d(SN1987A)=51.4\pm1.2\,kpc}\, ~~~~~~
\mathbf{(m-M)_\mathrm{SN1987A}=18.55\pm0.05}\, .$$

This value is very close to our original determination 
\cite{Panagia91} but is considerably more accurate. Actually, it could
still be an \emph{underestimate} to the true distance to SN\,1987A,
because if the ring is not perfectly circular, as hinted by the marginal
discrepancy between the inclinations determined from light curve fitting
($i\simeq 42^\circ$) and from the major to minor axis ratio ($i\simeq
44^\circ$), then the distance may have to be increased by as much as 2\%
\cite{Gould98}. 

Allowing for the difference of position of SN\,1987A relative to the LMC
barycenter \cite{vdMC01} the best estimate of the 
\textbf{distance to the center of mass of the LMC} is found to be 
$$\mathbf{d(LMC)=51.7\pm1.3\,kpc}\, ~~~~~~
\mathbf{(m-M)_\mathrm{LMC}=18.56\pm0.05\, .}$$ 

The error includes the uncertainties on the SN\,1987A distance ($\pm
1.2$\,kpc) as well as those on the depth toward SN\,1987A ($\pm0.2$\,kpc)
and the relative position of the LMC barycenter ($\pm 0.3$\,kpc).

\section {Discussion and Consequences for the Cosmological Distance Scale
\label{Sec.5}} 

In the literature one finds a number of determinations of SN\,1987A
distance which are all based on the analysis of UV line light curves
and \emph{HST} imaging but provide values that may be quite discrepant
with each other.  Table~\ref{Tab.1} summarizes most of the
``independent" analyses of such data, listing the authors (column~1),
the emission lines considered (column~2), the derived time of the onset
of the far side emission (column~3), the adopted/measured angular size
of the ring (column~4) and, finally, the resulting distance modulus
(column~5). 

\begin{table}
\centering
\caption{Summary of SN~1987A distance determinations based on UV line light curves and \emph{HST} imaging} 
\label{Tab.1}
\begin{tabular}{lllll}
\hline\noalign{\smallskip}
Authors                       & Emission Lines/Ions    &\multicolumn{1}{c}{$t_\mathrm{max}$}  
&\multicolumn{1}{c}{R$''$} &\multicolumn{1}{c}{(m-M)}	\\
& &\multicolumn{1}{c}{days} &\multicolumn{1}{c}{mas}  &\multicolumn{1}{c}{SN\,1987A}\\
\noalign{\smallskip}\hline\noalign{\smallskip}
\multicolumn{2}{l}{\emph{~~~Prelim.\ IUE reductions}}\\
\noalign{\smallskip}
Panagia et~al.\ 1991 
\cite{Panagia91}	& NIII], NIV], NV, CIII] & $413\pm24$ & $825\pm17$ & $18.55\pm0.13$\\
Gould 1995 
\cite{Gould95} 		        & NIII], NIV] 		 & $390\pm2$  & $858\pm11$ & $18.35\pm0.04$\\
\noalign{\smallskip}
\multicolumn{2}{l}{\emph{~~~Final IUE reductions}}\\
\noalign{\smallskip}
Sonneborn et al.\ 1997 
\cite{Sonneborn97} 	& NIII] 		             & $399\pm15$ & $858\pm11$ & $18.43\pm0.10$\\
Gould \& Uza 1998 
\cite{Gould98} 		& NIII], NIV] 		 & $378\pm5$  & $858\pm11$ & $18.37\pm0.04$\\
Panagia et~al.\ 2003 
\cite{Panagia03}        & NIII], CIII], OIII]  & $395\pm5$  & $808\pm17$ & $18.55\pm0.05$\\
\noalign{\smallskip}\hline
\end{tabular}
\normalsize
\end{table}

One sees immediately that most of the discrepancy can be attributed to
the different angular size adopted and/or to the selection of UV
emission lines that were employed to estimate the absolute size. 

In particular, the ``high" value of the angular size, 858\,mas, is the
\emph{average} of the sizes measured by   \cite{Plait95} on \emph{FOC} 
images taken mostly with  the [OIII] filter between August 1990 and
October 1993. Since the apparent  size of the ring increases with time,
such an average represent a gross  overestimate (about 6\%) of the ring
size at the time of the UV maximum  which leads to an underestimate of
the distance modulus of 0.13 magnitudes:  this effect \emph{alone}
accounts for most of the discrepancies. 

The second point to consider is the time of the far side emission
onset, $t_\mathrm{max}$. As mentioned in Sec.~\ref{Sec.2}, light
curves  of different ions give different values of $t_\mathrm{max}$.
This is due to  both measurement uncertainties and physical effects,
such as:
\begin{itemize}
\item different ions recombine at different rates; 
\item different lines react faster or slower to a
general temperature decline, i.e.\ cooling, depending on their
excitation potential;
\item the ring is made of a multitude of condensations with a wide range
of densities and temperatures, with the effect that intrinsic, and
possibly large fluctuations add on top of the measurement errors to
distort the average behaviour of light curves. 
\end{itemize}
To minimize these effect one has to combine the results of as many
light curves as possible but selecting only of ions with similar
characteristics, which is is what we have done in our study. 

The conclusion is that all apparent discrepancies can be explained in 
terms of less-than-perfect selections of the data to compare with each 
other.

Our geometric determination of the LMC distance modulus is in excellent
agreement with the recent determinations by Romaniello {\it et al.}
\cite{Romaniello00} that are based on  a  study of both Red Clump stars
and TRGB stars measured in multi-band HST images of SN~1987A  field.  
In particular, they obtained  $(m-M)_{RC}= 18.59\pm0.04\pm0.08$ and
$(m-M)_{TRGB} = 18.69 \pm 0.25 \pm 0.06$ (the quoted errors are the
statistical and systematic ones , respectively), whose weighted
average  is  $<(m-M)>_\mathrm{LMC~field}=18.60\pm0.04\pm0.08$. 

It is apparent that the true LMC distance modulus must be around 18.60
and that values lower than 18.48 and than higher 18.72 are to be
excluded with high confidence. 

The main consequence of our distance determination is that all Cepheid 
distances based on the canonical value of  18.50 for the LMC  ({\it
e.g.} \cite{Madore91}) should be increased by about 3\%. And, of
course,  all values of $H_0$ based directly or indirectly on Cepheid
distances should be reduced by the same amount.

In this light, I like to assess the consequences for the determination
of $H_0$ based on a Cepheid calibration of the peak brightness of
type~Ia  supernovae (SNIa) relatively nearby (up tp $\sim$25\,Mpc) and
comparison of Hubble diagrams of more distant SNIa. In a long term HST
project led by Sandage, Saha and Tammann, 9~SNIa in spiral galaxies
have been calibrated with Cepheid variables, resulting in average
absolute magnitudes for type~Ia supernovae $M_\mathrm{B}=-19.47\pm0.07$
and $M_\mathrm{V}=-19.46\pm0.06$ with the assumption of a LMC distance
modulus of 18.50 \cite{Saha01}.  Entering these values into the Hubble
diagram of more distant SNIa leads to values of the Hubble constant
around $H_0=61\pm6$\,km~s$^{-1}$\,Mpc$^{-1}$ for an adopted cosmological
model with $\Omega_M=0.3,~ \Omega_\Lambda=0.7$ \cite{Saha01}. 

As said above, the new LMC distance modulus would imply a reduced value
of the Hubble constant, by about -3\%.  However, one has to take  into
account reddening corrections for distant supernovae in the Hubble
diagram (this problem was partly bypassed in Sandage et~al.\  analysis
by considering a Hubble diagram that included only  SNIa affected  by
little reddening) whose effect may increase the value of $H_0$ by as
much as +7\% (see e.g.\,\cite{Hamuy96,Riess96}).  Combining  the two
competing effects in an approximate way results in a Hubble constant
of 
$$\mathbf{H_0=63\pm7\,\mathrm{km~s}^{-1}\,\mathrm{Mpc}^{-1}\, .}$$

Although it is obtained with a simplified analysis which can, and will
be refined, I regard this as a rather robust result that is not likely
to change much in the years to come, and that offers the pleasant
feature  of not violating any constraint posed by old stars and the
evolution of the Universe. 

%
%
\def \SAIT #1 #2 {Mem.\ Soc.\ Astron.\ It. \textbf{#1}, #2\ }
\def \MESS #1 #2 {The Messenger \textbf{#1}, #2\ }
\def \ASTRNACH #1 #2 {Astron.\ Nach. \textbf{#1}, #2\ }
\def \AAP #1 #2 {Astron.\ Astrophys. \textbf{#1}, #2\ }
\def \AAL #1 #2 {Astron.\ Astrophys.\ Lett. \textbf{#1}, L#2\ }
\def \AAR #1 #2 {Astron.\ Astrophys.\ Rev. \textbf{#1}, #2\ }
\def \AAS #1 #2 {Astron.\ Astrophys.\ Suppl.\ Ser. \textbf{#1}, #2\ }
\def \AJ #1 #2 {Astron.~J. \textbf{#1}, #2\ }
\def \ANNREV #1 #2 {Ann.\ Rev.\ Astron.\ Astrophys. \textbf{#1}, #2\ }
\def \APJ #1 #2 {Astrophys.~J. \textbf{#1}, #2\ }
\def \APJL #1 #2 {Astrophys.~J.\ Lett. \textbf{#1}, L#2\ }
\def \APJS #1 #2 {Astrophys.~J.\ Suppl. \textbf{#1}, #2\ }
\def \APSS #1 #2 {Astrophys.\ Space Sci. \textbf{#1}, #2\ }
\def \ASR #1 #2 {Adv.\ Space Res. \textbf{#1}, #2\ }
\def \BAIC #1 #2 {Bull.\ Astron.\ Inst.\ Czechosl. \textbf{#1}, #2\ }
\def \JSQRT #1 #2 {J.~Quant.\ Spectrosc.\ Radiat.\ Transfer \textbf{#1}, #2\ }
\def \MN #1 #2 {Mon.\ Not.~R.\ Astr.\ Soc. \textbf{#1}, #2\ }
\def \MEM #1 #2 {Mem.~R.\ Astr.\ Soc. \textbf{#1}, #2\ }
\def \PLR #1 #2 {Phys.\ Lett.\ Rev. \textbf{#1}, #2\ }
\def \PASJ #1 #2 {Publ.\ Astron.\ Soc.\ Japan \textbf{#1}, #2\ }
\def \PASP #1 #2 {Publ.\ Astr.\ Soc.\ Pacific \textbf{#1}, #2\ }
\def \NAT #1 #2 {Nature \textbf{#1}, #2\ }

%
%



\printindex
\end{document}